\documentclass[sigconf]{acmart}
\AtBeginDocument{%
  }

\setcopyright{acmlicensed}
\copyrightyear{2018}
\acmYear{2018}
\acmDOI{XXXXXXX.XXXXXXX}
\acmConference[Conference acronym 'XX]{Make sure to enter the correct
  conference title from your rights confirmation email}{June 03--05,
  2018}{Woodstock, NY}
\acmISBN{978-1-4503-XXXX-X/2018/06}

\usepackage{acronym}
\usepackage{amsmath}
\usepackage{multicol}
\usepackage{xspace}
\usepackage{url}
\usepackage{xcolor}      
\usepackage{enumitem}  
\usepackage{multirow}
\usepackage{balance}

\acrodef{AI}[AI]{Artificial Intelligence}
\acrodef{BoW}[BoW]{Bag-of-Words}
\acrodef{CAsT}[CAsT]{Conversational Assistance Track}
\acrodef{CS}[CS]{Conversational Search}
\acrodef{IR}[IR]{Information Retrieval}
\acrodef{LLM}[LLM]{Large Language Model}
\acrodef{RAG}[RAG]{Retrieval Augmented Generation}
\acrodef{RLHF}[RLHF]{Reinforcement Learning from Human Feedback}
\acrodef{NLP}[NLP]{Natural Language Processing}

\newcommand{\oursystem}{WalkRAG\xspace}

\usepackage[show]{chato-notes}

\newcommand{\chiaraP}[1]{{\color{blue}{#1}}}

\begin{document}

\title{Spatially-Enhanced Retrieval-Augmented Generation for Walkability and Urban Discovery}

\author{Maddalena Amendola, Chiara Pugliese}
\authornote{Both authors contributed equally to this research.}
\affiliation{%
  \institution{IIT-CNR}
  \city{Pisa}
  \country{Italy}
}

\author{Raffaele Perego, Chiara Renso}
\affiliation{%
  \institution{ISTI-CNR}
  \city{Pisa}
  \country{Italy}
}

\renewcommand{\shortauthors}{Amendola et al.}

\begin{abstract}
\acp{LLM} have become foundational tools in artificial intelligence, supporting a wide range of applications beyond traditional natural language processing, including urban systems and tourist recommendations. However, their tendency to hallucinate and their limitations in spatial retrieval and reasoning are well known, pointing to the need for novel solutions. Retrieval-augmented generation (RAG) has recently emerged as a promising way to enhance \acp{LLM} with accurate, domain-specific, and timely information. Spatial RAG extends this approach to tasks involving geographic understanding.
In this work, we introduce \oursystem, a spatial RAG-based framework with a conversational interface for recommending walkable urban itineraries. Users can request routes that meet specific spatial constraints and preferences while interactively retrieving information about the path and points of interest (POIs) along the way. Preliminary results show the effectiveness of combining information retrieval, spatial reasoning, and \acp{LLM} to support urban discovery.
\end{abstract}


\maketitle

\section{Introduction}\label{sec:intro}


\aclp{LLM} (LLMs) have become widely adopted across numerous applications, including urban environment applications, a field that increasingly demands intelligent systems that support sustainable, human-centric mobility 
and personalized tourism recommendations \cite{Wei24}. 
In this context, walking emerges not only as the most fundamental form of transportation but also as the most sustainable and beneficial to public health. Designing systems that promote walking, particularly through personalized tourist itinerary recommendations, is a broad and well-studied research area \cite{Contractor2021,10.1145/3696114}. However, traditional route recommendation systems often fall short in capturing this multidimensionality, relying heavily on shortest-path algorithms or limited user feedback.
To overcome these limitations, \ac{LLM}s have been recently explored in urban and mobility applications.
Despite their impressive capabilities, LLMs exhibit well-known limitations in tasks involving factual answering, spatial retrieval, and reasoning \cite{Li24,yu2025spatialragspatialretrievalaugmented}, underscoring the need for novel approaches to address these challenges. 

Retrieval-Augmented Generation (RAG, \cite{10.5555/3495724.3496517}) improves LLMs by grounding responses in content retrieved from controlled sources, helping mitigate hallucinations and improve factuality, especially useful in conversational applications. RAG operates in two stages: retrieval of relevant content based on the user query, and response generation using this information. The retrieval typically relies on specialized knowledge bases and dense neural models that map content and queries into a shared semantic vector space \cite{10.1145/3637870}.


Inspired by \cite{yu2025spatialragspatialretrievalaugmented}, we believe that a RAG mechanism can enhance LLM spatial reasoning and offer a promising direction for enhancing context-aware, personalized itinerary recommendations.  By coupling generative language models with spatial and contextual knowledge, spatial RAG systems can retrieve geographically grounded information,  generate personalized walking itineraries, and present them to the user in natural language. This approach enables dynamic and adaptive recommendations that align with user intents and local context.
In this short research paper, we thus investigate the following research question: 

\textit{RQ: how can we effectively exploit \ac{LLM}s to combine spatial reasoning and contextual urban knowledge to generate meaningful and walkable urban itineraries and support users in their fruition? }

 To answer this RQ, we design and evaluate \oursystem, a spatial RAG framework with a conversational interface for recommending walkable urban itineraries. Leveraging LLMs, users can ask in natural language for itineraries respecting specific spatial constraints and personal preferences, enabling personalized and context-aware route generation. Moreover, to further enhance the engagement and improve the walking experience, they can interactively retrieve information about the route or the points of interest (POIs) located along their walking paths.

Unlike prior work, \oursystem does not rely on a specific spatial database and instead focuses on open map data and dynamic conversation-based interaction. We specifically emphasize walkability, an increasingly central theme in urban studies \cite{Bartzokas21, Bartzokas23}, especially under the 15-minute city paradigm \cite{moreno2021introducing}.
While traditional walkability metrics (e.g., the Walkability Index \cite{Frank2010}) assess urban form using static spatial features, our approach complements them by dynamically tailoring itineraries based on user preferences and contextual environmental data encountered along the route.

\oursystem addresses the possible limitations of RAG systems \cite{MallenEtAl2023, RenEtAl2023}  by accurately retrieving relevant spatial or contextual content, feeding the LLM with targeted, query-driven information to enhance itinerary generation or to answer contextual user queries.
We validate \oursystem with reproducible experiments on a custom test dataset. Our experiments show that 
RAG significantly enhances factual and spatial accuracy and completeness: \oursystem consistently outperforms closed-book LLMs, which often suffer from spatial hallucinations and limited domain-specific knowledge.

The paper is structured in the following way. Section \ref{sec:framework} introduces our framework and describes its main components. The experimental settings and the \oursystem assessment are discussed in Section \ref{sec:experiments}. Finally, Section  \ref{sec:conclusions} presents conclusions and future work.

\section{The \oursystem framework}
\label{sec:framework}
\begin{figure*}[!ht]
    \centering
    \includegraphics[width=0.85\linewidth]{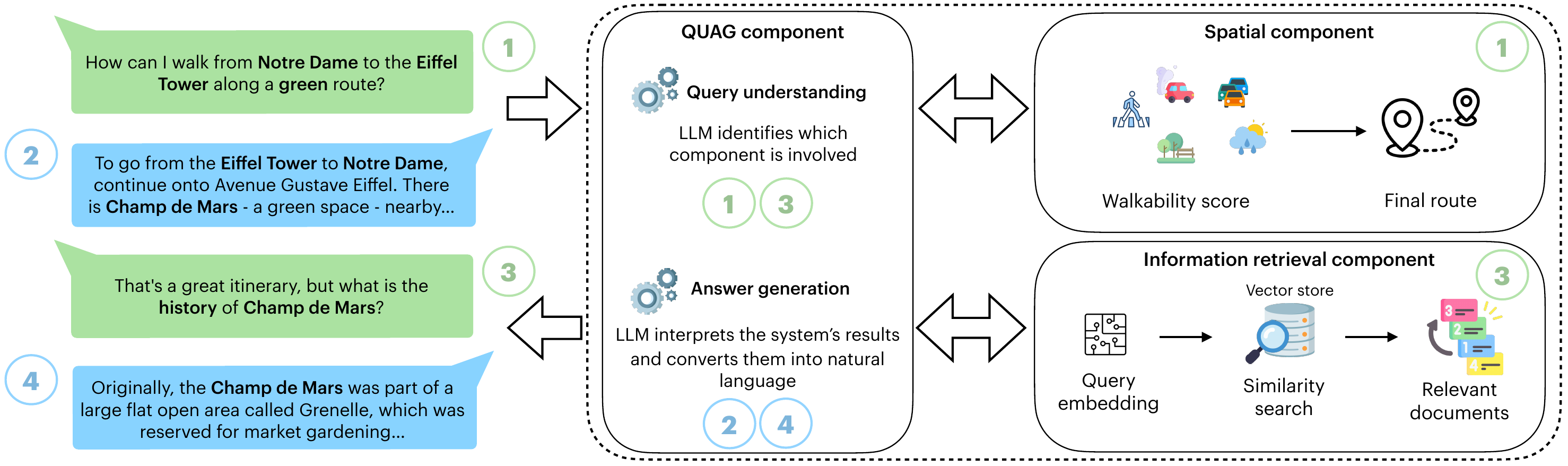}
    \caption{The \oursystem framework. The user query asking for a route from Notre Dame to the Eiffel Tower is redirected by the QUAG component to the spatial component for itinerary construction and walkability score computation (1). The answer returned to QUAG is interpreted by the LLM and returned to the user (2), who further interacts with the conversational system, asking for more details on Champs de Mars (3). QUAG now redirects the query to the Information Retrieval component, which retrieves the appropriate content from an index. The results retrieved are interpreted by the LLM and returned to the user (4). }
    \label{fig:framework}
\end{figure*}

\oursystem is a framework designed to enhance pedestrian experiences in urban environments by suggesting walkable routes toward specific destinations. It interacts with users through a friendly conversational interface, allowing them to easily access contextual information about locations and attractions along the suggested paths.
The framework comprises three main components, depicted in Figure \ref{fig:framework}: the Query Understanding and Answer Generation (QUAG) component, the Spatial component, and the Information Retrieval (IR) component. We detail their functionalities below.

\subsection{Query Understanding and Answer Generation}
This component encapsulates an \ac{LLM} and manages conversational interactions with users. Upon receiving an utterance, it determines whether the user’s information need pertains to: (i) a new itinerary suggestion taking into account walkability indicators and user preferences, 
or (ii) general information about attractions or points of interest related to a previously suggested itinerary.

In the first case, the query -- including the origin and destination -- is routed to the Spatial component of \oursystem, which generates a walkable itinerary (if any) between the specified points, enriched with auxiliary information that may be of interest to the user.
All other queries not involving itinerary requests are instead forwarded to the IR component.

In both scenarios, the information retrieved by the Spatial or IR components is used by QUAG to augment the generation capabilities of the integrated \ac{LLM}. QUAG then produces the final response based on the retrieved content.
When the Spatial component is involved, the response includes details about the suggested itinerary, emphasizing the walkability of the route and the points of interest encountered.
In the case of a general information request, the top-k results retrieved by the IR component from a knowledge repository are leveraged by QUAG to generate a contextually appropriate response, improving factual accuracy and reducing hallucinations.


\subsection{Spatial component}
Once the QUAG component determines that a user queries for itinerary information, it routes the request to the Spatial component. 
The primary objective of the Spatial component is to identify, among different alternatives, 
the most \textit{walkable} route between the specified origin and destination. To achieve this, inspired by \cite{Bartzokas21}, we incorporate a variety of indicators from heterogeneous data sources that can contribute to a high-quality walking experience: 
\begin{itemize}
    \item \textbf{Sidewalk/pedestrian footway availability}: sidewalks and pedestrian zones are fundamental for pedestrian safety and comfort, significantly enhancing route walkability.
    \item \textbf{Air pollution levels}: elevated pollution levels, often due to vehicular traffic, detract from the walking experience. 
    \item \textbf{Presence of green areas}: vegetation and green spaces improve the walking environment by providing shade and enhancing aesthetic and psychological comfort. 
    \item \textbf{Accessibility for individuals with disabilities}: we prioritized routes with curb ramps and smooth surfaces to ensure inclusivity and equitable access to accommodate individuals with mobility impairments.
\end{itemize}
We quantify the overall walkability of each candidate route as follows.
First, for each segment along a route, we count the occurrences of walkability indicators, capping the contribution of each segment at a maximum threshold $\tau$ to mitigate the impact of outliers. Second, for each indicator, we compute the average capped count per segment, denoted $c_i$, by dividing the total capped count by the number of segments. Third, we assign a user-defined weight $w_i$ to each indicator to reflect its relative importance, with the constraint $\sum_i w_i = 1$. In the absence of user input, uniform weighting is applied. Finally, the walkability score (WS) for the route is calculated as
$WS = \frac{\sum_i w_i \, c_i}{\tau}$.
Since the maximum possible value of $c_i$ is $\tau$, this normalization ensures that the walkability score ranges from 0 (completely unwalkable) to 1 (fully walkable).

In addition to standard walkability indicators, the component supports the enrichment of the route based on user preferences or contextual information. When such preferences are explicitly expressed in the query, the QUAG component identifies them and forwards the relevant parameters to this component. To incorporate these preferences, we generate a buffer around each alternative route and perform a spatial join to associate additional POIs or relevant features requested by the user.
If no explicit preferences are provided, we enrich routes with general tourist information, ensuring that the walking experience remains informative and engaging.



\subsection{Information Retrieval component}
The IR component integrates a neural indexing and search system.
Offline, documents from the knowledge base are encoded into dense vectors within a multidimensional latent space and stored in a vector index to enable efficient similarity-based retrieval.
At query time, each user query received by QUAG is similarly encoded into the same latent space. The resulting query representation is then compared to the indexed document vectors using an approximate nearest-neighbor search algorithm.
The top-k closest vectors retrieved from the index are considered the relevant context. The associated documents are returned to QUAG, which uses them to generate a grounded and contextually informed response.

\section{\oursystem assessment}
In this section, we describe the implementation of the three components of \oursystem, outline the experimental settings, and discuss the results along with the insights gained.
\label{sec:experiments}

\noindent \textbf{QUAG Component}.
The QUAG component is implemented in Python and manages the RAG-based conversational interface of \oursystem. It encapsulates 
the \textsf{Llama 3.1 8B} \ac{LLM} model\footnote{\url{https://huggingface.co/meta-llama/Llama-3.1-8B}} for query classification and retrieval-augmented answer generation.   

\noindent \textbf{Spatial Component}. After receiving from QUAG an itinerary suggestion request including the start and end points, we use \texttt{Nominatim} API\footnote{\url{https://nominatim.org/}} to identify the corresponding latitude and longitude coordinates. Then, we generate three alternative routes using GraphHopper API\footnote{\url{https://www.graphhopper.com/}}, which queries the footway road network from OpenStreetMap. 
The air quality index is retrieved by using the OpenWeatherMap’s API\footnote{\url{https://openweathermap.org/api}}. The remaining indicators and POI used to customize the route are retrieved with \texttt{OSMnx} library by filtering OpenStreetMap data using relevant tags such as landuse, natural, footway, wheelchair, and tourism. 
Regarding the walkability score, we set the threshold parameter to $\tau=5$ based on empirical observations. In the absence of user-defined preferences, we assign $0.25$ to each of the four indicators by default.
The output returned to QUAG is a JSON file containing information about the route with the highest walkability score, i.e., the routing instructions, the walkability score and indicators, and the list of POIs associated with each segment, together with their category and name.


\noindent \textbf{IR Component}.
As the retrieval corpus to support \oursystem information queries, we adopt the TREC Conversational Assistance Track (CAsT) 2019 and 2020 collection~\cite{dalton_cast_2019,dalton_cast_2020}. It includes three widely used datasets: TREC CAR (Complex Answer Retrieval), MS MARCO (MAchine Reading COmprehension)~\cite{NguyenRosenbergEtAl2016}, and Washington Post (WaPo). Together, these datasets comprise a total of 38.636.520 passages.
To retrieve relevant passages from these datasets in response to \oursystem queries, we leverage the FAISS vector search library \cite{johnson2019billion,douze2024faiss} and the \texttt{Snowflake}\footnote{\url{https://huggingface.co/Snowflake/snowflake-arctic-embed-l-v2.0}} bi-encoder, built on XLM-R Large and fine-tuned for retrieval tasks~\cite{yu2024arcticembed20multilingualretrieval}. The passages are encoded and indexed offline by computing their 1024-dimensional dense embeddings. At query time, the system encodes the incoming query and computes cosine similarity with the pre-computed embeddings to retrieve the most relevant passages.
Each query received from QUAG is processed by retrieving the top-3 passages from the index, which are then returned to QUAG for answer generation.

\noindent\textbf{\oursystem Dataset.}
For evaluation, we constructed a custom dataset composed of realistic walking and information-seeking queries within the city of Paris. The dataset consists of 10 distinct spatial requests, each paired with 3 follow-up information queries related to the route or nearby landmarks, for a total of 40 user queries. This design simulates a typical user interaction where a person first requests a walkable route and then engages in a conversation about places encountered along the way. 

\noindent \textbf{Reproducibility}.
The \oursystem dataset and the LLM instructions used are already available in our GitHub repository\footnote{\url{https://github.com/chiarap2/walkRAG/tree/main/dataset}}. The source code will be made available in the same repository upon publication.

\noindent
\textbf{Evaluation.} 
We compare the accuracy of the answers generated for the queries in our \oursystem dataset by two system configurations:
\begin{enumerate}
    \item \textbf{WalkRAG}, our open-book framework where spatial queries are answered based on route and environmental indicators by the spatial component, and information queries are grounded using the top-k passages retrieved by our IR component.
    \item \textbf{LLM-ClosedBook (LLM-CB)}, a baseline configuration where the LLM model (Llama 3.1 8B) is used in isolation without any route enrichment or external retrieval augmentation.
\end{enumerate}

\subsection{Results and discussion}

In this section, we present the results of our evaluation using the queries of the \oursystem dataset. We focus separately on the two types of user interactions addressed: \textit{spatial} and \textit{information} requests. 
For the 10 spatial requests, we evaluate whether the LLM can generate coherent routes, assess their walkability, and suggest relevant urban entities based on user preferences. For the 30 Information requests, we assess instead the model's ability to provide accurate and contextually appropriate answers to general-purpose queries about urban entities encountered along the route. Finally, we assess whether \oursystem mitigates the limitations of the LLM-CB baseline in both spatial and informational tasks by leveraging proper contextual and geographical knowledge. The summary of the results achieved is reported in Table \ref{tab:results}.
\begin{table}[ht]
    \caption{Summary of the evaluation results}
    \label{tab:results}
    \centering
    \footnotesize
    \resizebox{0.98\columnwidth}{!}{%
    \begin{tabular}{lcccc}
    \toprule
    \textbf{System}                          & \textbf{Query type}  &  \textbf{Correct}  & \textbf{Partially correct}  & \textbf{Incorrect}\\ \midrule
    \multirow{2}{*}{LLM-CB}         & Spatial  & 0   & 0 & 10\\
                                    & Information & 12 & 11 & 7\\
    \midrule
    \multirow{2}{*}{\oursystem}     & Spatial  &  4  &  6 & 0\\
                                    & Information & 20 & 5 & 5 \\
    \bottomrule
    \end{tabular}
}
\end{table}

\noindent\textbf{Query understanding.}
In our experiments, the integrated LLM allowed the QUAG component to correctly classify all 40 queries and route them to the appropriate spatial or IR component.

\begin{figure}[!h]
    \centering
    \includegraphics[width=0.85\linewidth]{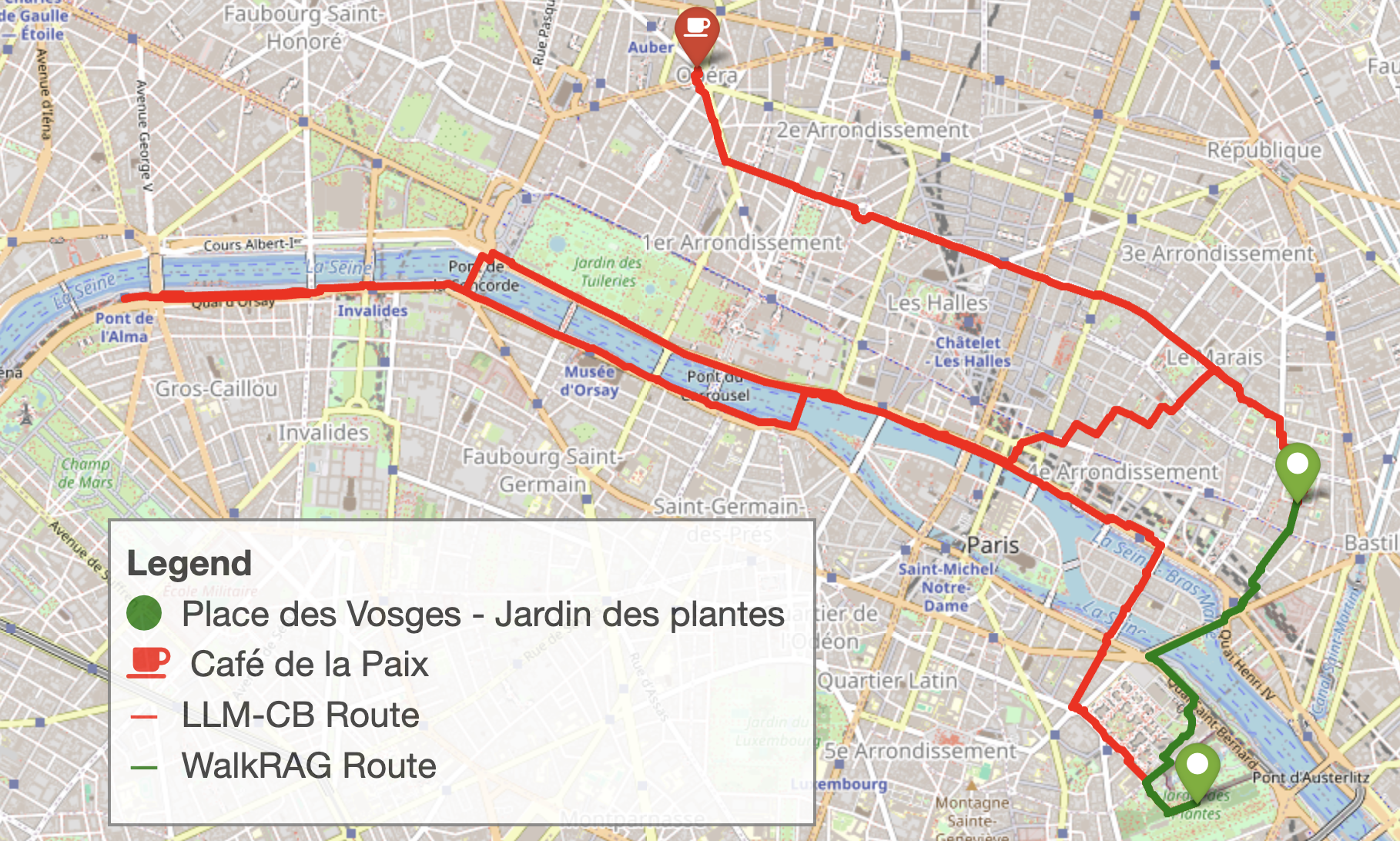}
    \caption{LLM‑CB and \oursystem routes for the third spatial query of the dataset.}
    \label{fig:mapR3}
\end{figure}

\noindent\textbf{Spatial requests.} 
To evaluate the effectiveness of the proposed Spatial RAG mechanism, we analyzed the responses to the 10 route-based queries of our dataset. A route is considered correct if it leads the user from the specified origin to the destination in a continuous way, taking into account walkability indicators and any user-defined preferences. 
LLM-CB failed all 10 queries. Its responses consistently contained hallucinations, such as suggesting directions that exhibited significant jumps from the intended path (ranging from 1.7 km to 8.6 km), looping instructions, and poor spatial awareness (e.g., confusion between left and right). Furthermore, it frequently recommended POIs located far from the actual route, such as for the third spatial query of the dataset, in which LLM-CB suggested to visit \textit{Café de la Paix}, which lies 3.9 km from the \textit{Jardin des Plantes} destination. This example is shown in Figure \ref{fig:mapR3}. 
By contrast, \oursystem returned 4 fully correct routes and 6 partially correct ones. Partial correctness was defined by minor omissions in navigation steps. 
To achieve these results, we experimented with various instruction formulations. Less structured prompts produced responses more fluent but of lower accuracy, whereas the more schematic prompt resulted in higher accuracy but reduced textual fluency. Importantly, \oursystem did not produce hallucinations. In most of the partially correct responses, the missing steps were duplicates of earlier instructions (e.g., repeated turns or identical POIs), and only one case omitted a "continue" instruction.

In terms of walkability, \oursystem accurately identified both of the poorly walkable routes in the dataset (the fourth and the fifth), incorporating user preferences into its assessment. LLM-CB, on the other hand, correctly flagged only one of these, and only due to an overestimation caused by an erroneous 10-km route. For the other, it recommended public transportation but directed users to a station roughly 5 km away from the intended destination. Notably, across all LLM-CB responses, the initial and final instructions were typically aligned with the start and end points, while the intermediate steps exhibited substantial disorientation.

\noindent\textbf{Information requests.}
To evaluate the effectiveness of the IR-based RAG mechanism in \oursystem, we analyzed the system's responses to the 30 information queries included in our dataset. Each response was manually labeled as correct, incorrect, or partially correct (imprecise). A response was deemed \textit{partially correct} when it conveyed the correct general information but included factual inaccuracies or lacked specificity.
For instance, a typical partially correct answer occurred when responding to the question \textit{“What are the most important hotels in Paris?”}. In this case, the model listed several relevant hotels, but also included the \textit{Hotel du Petit Bourbon}, a historical building that was demolished in the 17th century. While the answer captures the intended topic (notable hotels), it failed to distinguish between current and historical relevance.

The overall results achieved clearly highlight the benefit of RAG:
    WalkRAG returned 20 correct answers, 5 partially correct, and 5 incorrect ones. Notably, in 3 of the incorrect answers, the system failed to retrieve relevant information from the indexed collection, which prevented the LLM from generating a grounded response.
    LLM-CB, by contrast, produced 12 correct answers, 11 partially correct, and 7 incorrect ones.

\section{Conclusion and future work}
\label{sec:conclusions}

The potential of LLMs in the urban domain is considerable. However, they exhibit well-documented limitations in spatial reasoning. To address this, we introduce \oursystem, a spatially-enhanced retrieval-augmented framework equipped with a conversational interface for recommending personalized, walkable urban itineraries. \oursystem enables users to define spatial constraints and personal preferences, and retrieve contextual information about attractions along the route.
Our experiments show that retrieval-augmented generation significantly enhances factual accuracy and completeness. While \oursystem may falter when retrieval lacks sufficient context, it consistently outperforms closed-book LLMs, which suffer from spatial/factual hallucinations and limited contextual knowledge.
Findings highlight how LLMs alone struggle to generate coherent, walkable itineraries or to suggest urban elements in an exploratory context. They also underperform on general-purpose queries on not highly popular topics. 
These preliminary results open avenues for future work:
(1) assessing the impact of different LLM model sizes and architectures on RAG performance;
(2) enhancing spatial reasoning through richer geographic operations for walkability and the use of routing algorithms; and
(3) studying how LLMs process structured route data -- particularly their tendency to omit repeated instructions -- potentially via improved spatial encoding in the RAG pipeline.

\begin{acks}
This work was supported by PNRR - M4C2 - Investimento 1.3, Partenariato Esteso PE00000013 - “FAIR - Future Artificial Intelligence Research” - Spoke 1 ”Human-centered AI”, funded by the European Commission under the NextGeneration EU programme and by the European Union under the Italian National Recovery and Resilience Plan (NRRP) of NextGenerationEU, partnership on “Telecommunications of the Future” (PE00000001 - program “RESTART”). This research has been partially funded by the European Union’s Horizon Europe research and innovation program EFRA (Grant Agreement Number 101093026). Views and opinions expressed are however those of the authors only and do not necessarily reflect those of the European Union or European Commission-EU. Neither the European Union nor the granting authority can be held responsible for them.
\end{acks}

\balance
\bibliographystyle{ACM-Reference-Format}
\bibliography{biblio}



\end{document}